\documentclass[twocolumn,aps,prl,superscriptaddress,longbibliography]{revtex4-1}
\usepackage[colorlinks,bookmarks=false,citecolor=blue,linkcolor=red,urlcolor=blue]{hyperref}
\usepackage{verbatim}
\usepackage{amsmath,amssymb,bm,braket,float,mathtools}
\usepackage{graphicx,xfrac,appendix}
\usepackage{CJK}
\usepackage[english]{babel}
\makeatletter

\usepackage{epstopdf}

\makeatother

\begin{document}
\title{Topological Floquet-bands in a circularly shaken dice lattice}

\author{Shujie Cheng}
\affiliation{Department of Physics, Zhejiang Normal University, Jinhua 321004, China}
\author{Gao Xianlong}
\affiliation{Department of Physics, Zhejiang Normal University, Jinhua 321004, China}
\date{\today}

\begin{abstract}
The hoppings of non-interacting particles in the optical dice lattice result in the gapless dispersions in the
band structure formed by the three lowest minibands. In our research, we find that once a periodic driving force is applied to
this optical dice lattice, the original spectral characteristics could be changed, forming three gapped quasi-energy
bands in the quasi-energy Brillouin zone. The topological phase diagram containing the Chern number of the
lowest quasi-energy band shows that when the hopping strengths of the nearest-neighboring hoppings are isotropic,
the system persists in the topologically non-trivial phases with Chern number $C=2$ within a wide range of the driving strength.
Accompanied by the anisotropic nearest-neighboring hopping strengths, a topological phase transition occurs,
making Chern number change from $C=2$ to $C=1$. This transition is further verified by our analytical method. Our theoretical work
implies that it is feasible to realize the non-trivially topological characteristics of optical dice lattices by applying the periodic shaking, and
that topological phase transition can be observed by independently tuning the strength of a type of nearest-neighbor hopping.
 \end{abstract}

\maketitle

\section{Introduction}
The emergence of the non-trivial topology in the band insulators is relevant to the properties of their band structure.
The resulting topological band insulators \cite{TI_1,TI_2} are not only classified by the symmetries but also are found 
to be immune to the inhomogeneous perturbations because of their preserved symmetries \cite{Classfi_1,Classfi_2,Classfi_3}. 
The topological band theory proposed by Thouless, Kohmoto, Nightingale, and den Nijs (TKNN) \cite{TKNN} tells that
if a band insulator is capable of changing between the topological trivial and non-trivial phase, there shall exist tunable
band inversion points (or say Dirac points), at which bands either are non-degenerate or degenerate. Accordingly, the 
crucial factor to engineer topological band structures is controlling the degeneracies of the bands at these inversion 
points \cite{BSE}, which is also a crucial factor to prepare a band insulator with quantum anomalous Hall effect \cite{QAHE_1,QAHE_2}. 
Nevertheless, in practice, it remains a challenge to realize the flexible control on the band inversion points
in experiments \cite{challenge_1,challenge_2}.

Alternatively, the control of band characteristic can be realized by the Floquet engineering which offer a new way to 
study the dynamical properties of topological matters \cite{BSE,Eckardt}. Due to the periodic driving, the intrinsic 
trivial characteristic of the band structures of the static system changes, and non-trivial Floquet quasi-energy bands 
occur. For decade, the scheme of Floquet band engineering has been employed in solid-state materials \cite{material_1,material_1_1,material_1_2,material_2,material_3,material_4,material_5,material_6}, photonic 
systems \cite{photonic_1,photonic_2}, and the ultracold atoms systems \cite{BSE,Eckardt,coldatom_1,coldatom_2,coldatom_3,coldatom_4,coldatom_5,coldatom_6,coldatom_6_1,coldatom_7,coldatom_8}.
Recently, an experimentally and theoretically investigation \cite{Esslinger} on the feasibility to individually control
the quasi-energy bands coupling and decoupling at band inversion points is carried out in a one-dimensional lattice
by tuning the strength of the periodic shaking. Motivated by this Floquet engineering, we want to study whether it is
possible to decouple the intrinsic gapless bands and form gaped quasi-energy band structures with non-trivial topology
in a driven two-dimensional dice optical lattice by only tuning the shaking strength.

We note that recent research shows that if circular frequency light \cite{material_1,material_1_1,material_1_2} is applied
to a class of $\alpha$-$\mathcal{T}_{3}$ lattice, a non-trivial topology can be induced 
and the topological phase transition is dominated by the parameter $\alpha$ \cite{Bashab}, 
which controls all types of nearest-neighbor hoppings in the system. 
The dice lattice we considered \cite{dice_1,dice_2,CN_cal_1,dice_laser_1,dice_laser_2,dice_laser_3} 
is topologically equivalent to the $\alpha$-$\mathcal{T}_{3}$ lattice \cite{Bashab,Biswas}. The hoppings between each type of 
nearest-neighbor sublattice sites in our dice system are independently assisted by the Raman lasers \cite{CN_cal_1,dice_laser_1,dice_laser_2,dice_laser_3}. 
Except for the circular frequency light, periodic shaking is found to be an effective way to induce topologically non-trivial 
band structures \cite{coldatom_4,Esslinger}, which has been successfully implemented in the Haldane model \cite{coldatom_4,Haldanelattice}. 
In our research, from the perspectives of the real space and quasi-momentum space, we will explore some new applications
of the periodic shaking method in designing topologically nontrivial band structures and study whether the individual control of the
nearest-neighbor hopping can realize the topological transition in the shaken two-dimensional dice system.

\section{Dice lattice and the Floquet engineering}
 \begin{figure}[htp]
  \centering
  \includegraphics[width=0.5\textwidth]{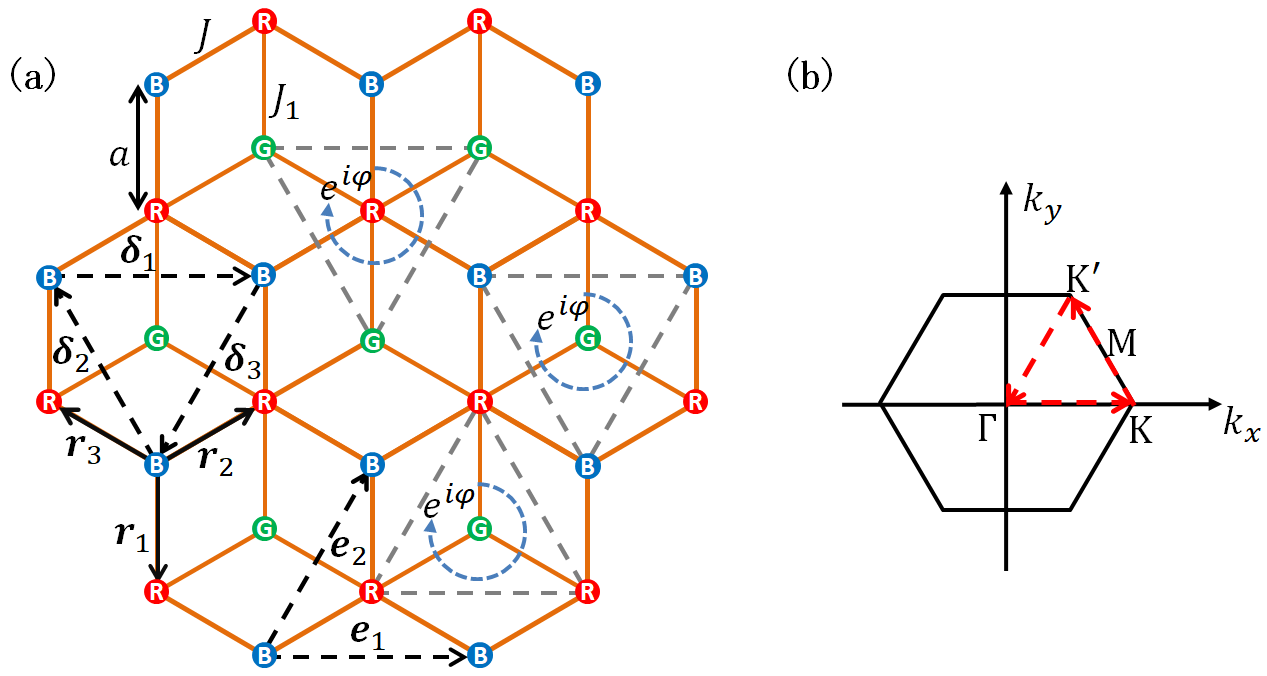}
  \caption{(Color Online) (a) Sketch of the dice lattice. In each unit cell constructed by the primitive lattice vectors $\bm{e}_{1}$ and $\bm{e}_{2}$, there are three
  types of sublattices R (red dot), B (blue dot), and G (green dot).  Nearest-neighbor sites are connected by vectors $\bm{r}_{a}$ ($s=1,2,3$) and next-nearest-neighbor
  sites are connected by vectors $\bm{\delta}_{s}$ ($s=1,2,3$). In the initial case, particles hop between neighboring $R$ and $B$ sites with hopping strength $J$
  and $R$ and $G$ with $J_{1}$. The induced next-nearest-neighbor hopping between the same R/B/G sublattice sites is
  accompanied by a phase $e^{i\varphi}$ ($\varphi=\frac{\pi}{2}$). (b) The first Brillouin zone obtained according to the primitive lattice vectors.
  $\Gamma$-K-K'-$\Gamma$ is high-symmetry path shown by red dashed arrows.}\label{f1}
 \end{figure}

We note that a retro-reflected laser can be used to create one-dimensional periodic potential wells for ultracold atoms,
and the resulting phenomena can be interpreted by a tight-binding model with multiple bands \cite{Esslinger}. Following this way,
a two-dimensional three-band dice optical lattice \cite{CN_cal_1,dice_laser_1,dice_laser_2,dice_laser_3} can be created for spinless and
non-interacting ultracold fermionic atoms by employing three retro-reflected lasers \cite{dice_laser_1}, as illustrated in the
schematic Fig.~\ref{f1}(a). Intuitively, the system preserves the discrete translational symmetry. In each unit cell braided by two primitive
lattice vectors, there are three types of sublattices shown by $R$, $B$, and $G$. Taking a similar strategy as that in Ref. \cite{Esslinger},
we theoretically study the system by using the tight-binding approximation as well. The generalized tight-binding Hamiltonian is initially given by,

\begin{equation}
 \begin{aligned}
 \hat{H}_{\rm ini}&=\sum_{\langle \bm{R}_j,\bm{B}_{j'}\rangle}J\left(\hat{c}^\dag_{\bm{R}_j}\hat{c}_{\bm{B}_{j'}}+H.c.\right)\\
 &+\sum_{\langle \bm{G}_{j},\bm{R}_{j'}\rangle}J_1\left(\hat{c}^\dag_{\bm{G}_j}\hat{c}_{\bm{R}_{j'}}+H.c.\right).
 \end{aligned}
\end{equation}
$\hat{H}_{\rm ini}$ describes the hoppings between the nearest-neighbor sites with $\bm{\alpha}_{j}$ ($\bm{\alpha} \in \{\bm{R, G, B}\}$)
being the coordinate of the lattice site and $j$ being the site index. $J$ is the hopping strength between the nearest-neighbor $R$ and $B$
sites, and $J_1$ is the one between the nearest-neighbor $G$ and $R$ sites. The summation are on all the nearest-neighbor
relations $\langle \bm{\alpha}_j,\bm{\alpha'}_{j'}\rangle$. $J$ is taken as the unit of energy, and we consider two systems
with the isotropic case for $J_{1}=J$ and the anisotropic one for $J_{1}\neq J$.

Our Floquet band engineering is to apply an anisotropic time-dependent shaking force $\bm{F}(t)$ on the initial lattice platform
with $F\cos(\omega t)$ in $\bm{e}_{x}$ direction and $-F\sin(\omega t)$ in the $\bm{e}_{y}$ direction, like the shaking of the
Haldane lattice \cite{coldatom_4}, where $F$ and $\omega$ indicate the strength and the frequency of the shaking force, respectively.
In fact, circular driving is initially proposed by Oka and Aoki. In Ref. \cite{material_1}, they use the circular frequency light to induce
the non-trivial topology in the Graphene system. Circular shaking is a natural development of circular driving in the mechanical branch.
In our research, we want to explore some new applications of the periodic shaking method in designing topologically nontrivial band structures.
The driving force is described by the time-dependent on-site potential $\hat{H}_{\rm dri}$,
\begin{equation}
\hat{H}_{\rm dri}=\sum_{\alpha_{j}}V(\bm{\alpha}_{j},t)\hat{n}_{\bm{\alpha}_{j}},
\end{equation}
where $V(\bm{\alpha}_{j},t)=-\bm{\alpha}_{j}\cdot \bm{F}(t)$ and $\hat{n}_{\bm{\alpha}_{j}}=\hat{c}^\dag_{\bm{\alpha}_j}\hat{c}_{\bm{\alpha}_{j}}$.

\section{Effective Hamiltonian}
The attitude of our Floquet band engineering is to theoretically investigate the feasibility that inducing gapped band structures and preparing non-trivial
topological phases can be realized by only tuning the driving strength. Therefore, the driving strength $F$ can ranges from zero to a finite value. We derive the effective Hamiltonian by the Floquet analysis \cite{Eckardt,Floquet_1,Floquet_2}. Transforming
the total Hamiltonian to the rotating frame, we have the gauge-transformed Hamiltonian $\hat{H}_{\rm rot}$ as (see the derivation in Appendix \ref{A1})
\begin{equation}\label{H_rot}
\begin{aligned}
\hat{H}_{\rm rot}&=\hat{U}^\dag(t)\left[\hat{H}_{\rm ini}+\hat{H}_{\rm dri}\right]\hat{U}(t)-i\hbar\hat{U}^\dag(t)\frac{d}{dt}\hat{U}(t)\\
&=\hat{U}^\dag(t)\hat{H}_{\rm ini}\hat{U}(t),
\end{aligned}
\end{equation}
where $\hat{U}(t)$ is the time-dependent gauge transformation operator and $\hat{U}(t)=\exp\left(-\frac{i}{\hbar}\sum_{\alpha_{j}}
 \int^{t}_{0}V(\bm{r}_{\alpha_{j}},t)dt\cdot\hat{n}_{\alpha_{j}}\right)$.

 \begin{figure}[htp]
  \centering
  \includegraphics[width=0.45\textwidth]{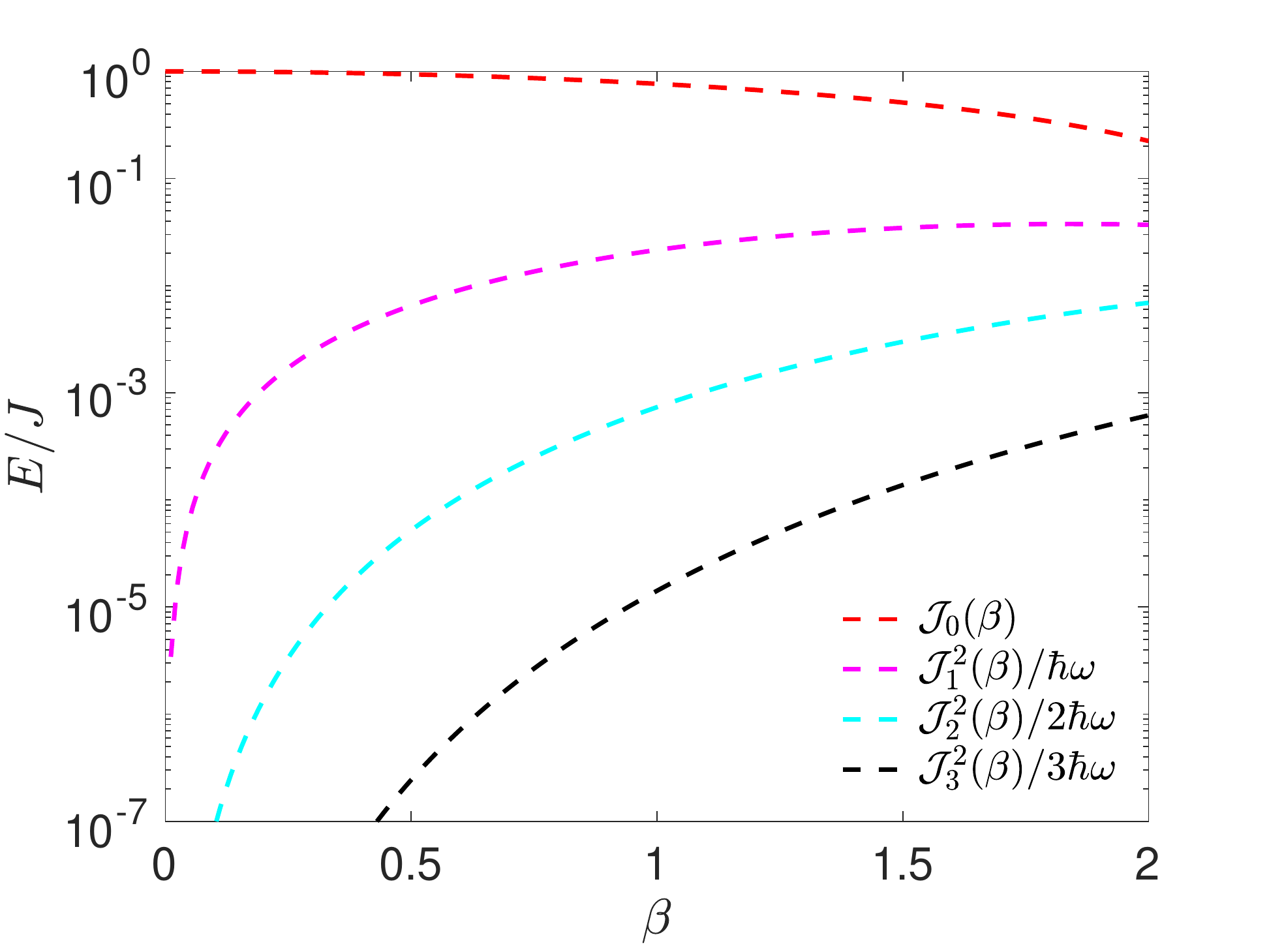}\\
  \caption{(Color Online) Components of the hopping strengths as a function of the dimensionless
  driving strength $\beta=Fa/\hbar\omega$. The term $\mathcal{J}_{0}(\beta)$ contributes to the
  nearest-neighbor hopping strengths and the terms $\mathcal{J}^2_{p}(\beta)/p\hbar\omega$ contribute to
  the next-nearest-neighbor hopping strengths. $\mathcal{J}_{p=0,1,2,3}$($\beta$) is the $p$th-order Bessel
  function. The involved parameter is $\hbar\omega=9J$.}\label{f2}
 \end{figure}

 According to the method discussed in Refs. \cite{Eckardt}, $\hat{H}_{\rm rot}$ can be rewritten as
 \begin{equation}\label{H_m}
 \hat{H}_{\rm rot}=\hat{\mathcal{H}}_{0}+\sum_{m=1}^{\infty}\left[\hat{\mathcal{H}}_{m} e^{i m \omega t}+\hat{\mathcal{H}}_{-m} e^{-i m \omega t}\right],
 \end{equation}
 where $\hat{\mathcal{H}}_{\pm m}$ is the Fourier  component of $\hat{H}_{\rm rot}$ (see the derivation in Appendix \ref{A2}). With these components,
 the effective Hamiltonian $\hat{H}_{\rm eff}$ is given by
 \begin{equation}\label{eq5}
 \hat{H}_{\rm eff}=\hat{\mathcal{H}}_{0}+ \sum^{\infty}_{m=1} \frac{1}{m\hbar\omega}\left[\hat{\mathcal{H}}_{m}, \hat{\mathcal{H}}_{-m}\right]+\mathcal{O}(1/\omega^2),
 \end{equation}
where the first-order approximation to the effective Hamiltonian is considered.

 In the following analyses, we take a fast and approximate driving frequency with $\omega=9J/\hbar$ as an example,
 which can not only avoid the multi-photon couplings to higher bands and the multiple bands mixing within the low-energy
 subspace \cite{coldatom_8,fre_1,fre_2,fre_3,fre_4,fre_5} but also offer finite band gaps. The size of the
 gap is determined by the effective tunneling strengths which depend on the driving frequency. For the experiments
 with cold atoms, large band gaps compared to temperature must be achieved to establish a topological state \cite{coldatom_4,coldatom_8}.
 Besides, in our analyses, the maximal strength of the shaking force is limited to twice of the frequency,
 i.e., $F_{max}a=2\hbar\omega$ ($\hbar=1$). Figure \ref{f2} presents the terms $\mathcal{J}_{0}(\beta)$ and
 $\mathcal{J}^2_{p}(\beta)/p\hbar\omega$ ($\mathcal{J}_{p=0,1,2,3}$($\beta$) is the $p$th-order Bessel function) as
 a function of dimensionless driving strength $\beta=Fa/\hbar\omega$, contributing to the nearest-neighbor hopping
 strengths and the next-nearest-neighbor hopping strengths, respectively. Intuitively, the high order term $\mathcal{J}^2_{3}(\beta)/3\hbar\omega$
 is always less than $10^{-3}$ at each given $\beta$, so it reasonable to truncate the $\hat{H}_{\rm eff}$ until $m=2$.
 Finally, the effective Hamiltonian is obtained as

 \begin{equation}
 \begin{aligned}
 \hat{H}_{\rm eff}&=\hat{\mathcal{H}}_{0}+\sum_{m=1,2}\left[\hat{\mathcal{H}}_{m}, \hat{\mathcal{H}}_{-m}\right]+\mathcal{O}(m\geq 3) \\
 &=\sum_{\langle \bm{R}_{j},\bm{B}_{j'}\rangle}t_{rb}\hat{c}^\dag_{\bm{R}_j}\hat{c}_{\bm{B}_{j'}}+\sum_{\langle \bm{G}_{j},\bm{R}_{j'}\rangle}t_{gr}\hat{c}^\dag_{\bm{G}_j}\hat{c}_{\bm{R}_{j'}} \\
 &+\sum_{\ll \bm{R}_{j},\bm{R}_{j'}\gg}t_{rr}\hat{c}^\dag_{\bm{R}_j}\hat{c}_{\bm{R}_{j'}}+\sum_{\ll \bm{B}_{j},\bm{B}_{j'}\gg}t_{bb}\hat{c}^\dag_{\bm{B}_j}\hat{c}_{\bm{B}_{j'}}\\
 &+\sum_{\ll \bm{G}_{j},\bm{G}_{j'}\gg}t_{gg}\hat{c}^\dag_{\bm{G}_j}\hat{c}_{\bm{G}_{j'}}+h.c,
 \end{aligned}
 \end{equation}
 where $\ll \cdots \gg$ indicates the next-nearest-neighbor hoppings between the sublattice sites of the same type, and the hopping parameters are
 \begin{equation}
 \begin{aligned}
 &t_{rb}=J\mathcal{J}_{0}\left(\beta\right), \\
 &t_{gr}=J_{1}\mathcal{J}_{0}\left(\beta\right), \\
 &t_{rr}=\frac{\sqrt{3}}{2}e^{i\varphi}\left(\frac{J^2-J^2_{1}}{\hbar\omega}\right)\left[\mathcal{J}^{2}_{1}\left(\beta\right)-\frac{1}{2}\mathcal{J}^2_{2}\left(\beta\right)\right], \\
 &t_{bb}=\frac{\sqrt{3}J^2}{2\hbar\omega}e^{i\varphi}\left[\mathcal{J}^{2}_{1}\left(\beta\right)-\frac{1}{2}\mathcal{J}^2_{2}\left(\beta\right)\right], \\
 &t_{gg}=-\frac{\sqrt{3}J^2_1}{2\hbar\omega}e^{i\varphi}\left[\mathcal{J}^{2}_{1}\left(\beta\right)-\frac{1}{2}\mathcal{J}^2_{2}\left(\beta\right)\right],
 \end{aligned}
 \end{equation}
 where $\varphi=\pi/2$.

Having considered that the dice system preserves the translational symmetry, we can perform a SU(3) mapping \cite{SU3} to
transform the real-space $\hat{H}_{\rm eff}$ into the quasi-momentum space.
Based on the basis $\left(\hat{c}_{\bm{k},R},\hat{c}_{\bm{k},B},\hat{c}_{\bm{k},G}\right)^{\rm T}$ where
$\hat{c}_{\bm{k},\alpha}=\frac{1}{\sqrt{N}}\sum_{\bm{\alpha}_j}e^{-i\bm{k}\cdot\bm{\alpha}_j}\hat{c}_{\bm{\alpha}_j}$ is the Fourier operation,
the effective Bloch Hamiltonian is obtained as
\begin{equation}\label{H_eff}
\hat{H}_{\rm eff}=\left(
\begin{array}{ccc}
d_{3}+d_{8} & d_{1}-id_{2} & d_{4}-id_{5} \\
d_{1}+id_{2} & -d_{3}+d_{8} & 0\\
d_{4}+id_{5} & 0 & -2d_{8}
\end{array}
\right),
\end{equation}
in which the matrix elements are
\begin{equation}\label{elements}
 \begin{aligned}
 d_1&=t_{rb}\sum_{s}\cos\left(\mathbf{k}\cdot\bm{r}_s\right),~~d_2=t_{rb}\sum_{s}\sin\left(\mathbf{k}\cdot\bm{r}_s\right),\\
 d_3&=-(|t_{gg}|+2|t_{rr}|)\sum_{s}\sin(\bm{k}\cdot\bm{\delta}_s),\\
 d_4&=t_{gr}\sum_{s}\cos\left(\mathbf{k}\cdot\bm{r}_s\right),~~d_5=-t_{gr}\sum_{s}\sin\left(\mathbf{k}\cdot\bm{r}_s\right),\\
  d_8&=|t_{gg}|\sum_{s}\sin(\bm{k}\cdot\bm{\delta}_s),
 \end{aligned}
 \end{equation}
  where, the bond length has been set as $a=1$, and the six vectors $\bm{r}_s$ and $\bm{\delta}_s$ ($s=1,2,3$) shown
  in Fig.~\ref{f1}(a) are
 \begin{equation}
 \begin{aligned}
 \bm{r}_1&=\binom{0}{-1},\quad\bm{r}_2=\frac{1}{2}\binom{\sqrt{3}}{1},
 \quad\bm{r}_3=\frac{1}{2}\binom{-\sqrt{3}}{1},\\
 \bm{\delta}_1&=\binom{\sqrt{3}}{0},\quad\bm{\delta}_2=\frac{1}{2}\binom{-\sqrt{3}}{3},
 \quad\bm{\delta}_3=-\frac{1}{2}\binom{\sqrt{3}}{3}.
 \end{aligned}
 \end{equation}

\begin{figure}[htp]
  \centering
  \includegraphics[width=0.45\textwidth]{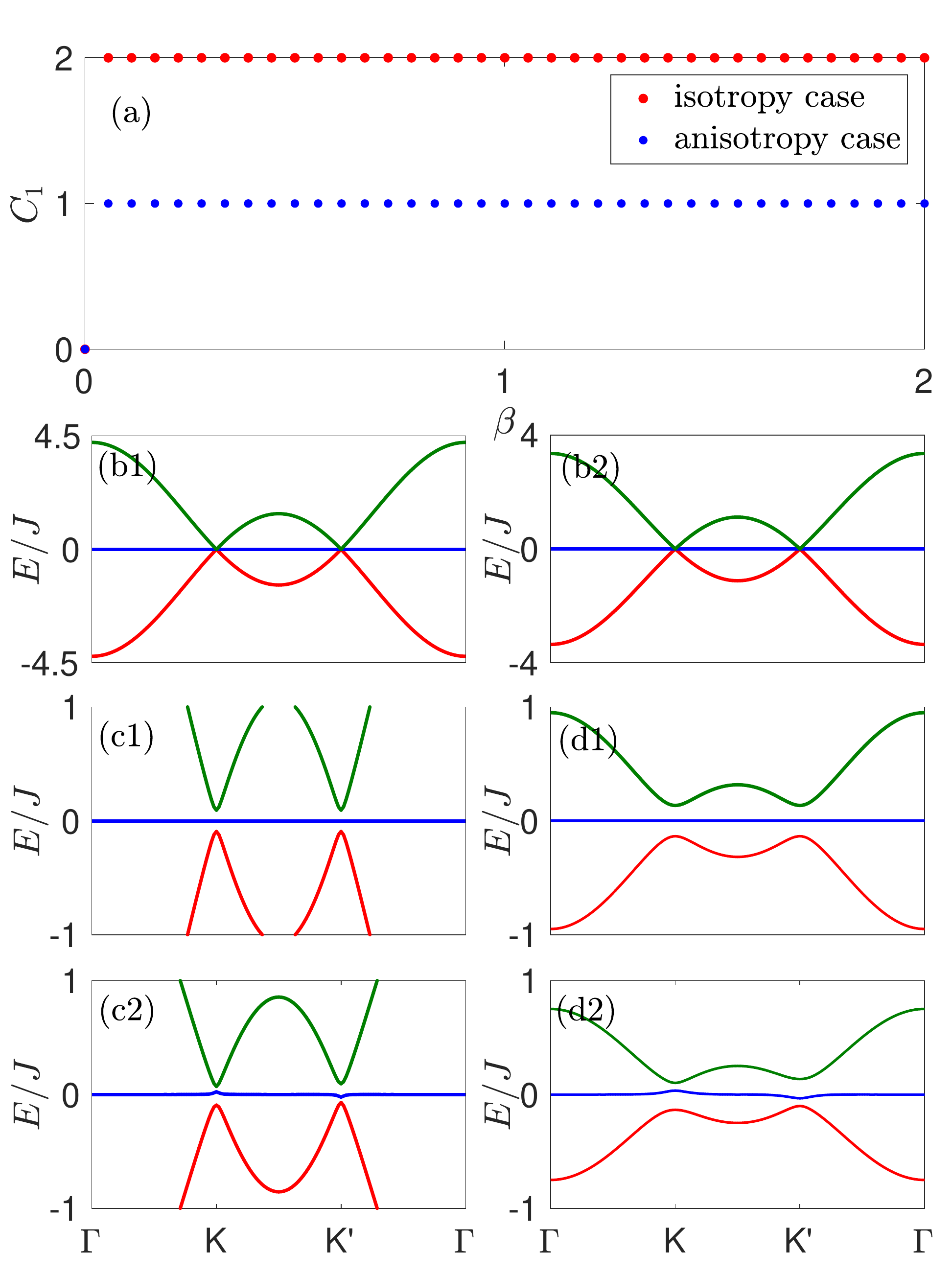}\\
  \caption{(Color Online)
  (a) Topological phase diagram contains the Chern number of the lowest quasi-energy band $C_{1}$ as a
  function of $\beta$ with red dots for the isotropic case and blue dots for the anisotropic case. The Chern
  numbers of the middle band $C_{2}$ for the two cases are equal to zero, which are not shown.  (b1) and (b2)
  Gapless dispersions without driving ($\beta=0$), corresponding to the isotropic case and anisotropic one,
  respectively. (c1) and (d1) Driving induced gapped dispersions under the isotropic case with parameters
  $\beta=1$ and $\beta=2$, respectively. (c2) and (d2) Driving induced gapped dispersions under the aniisotropic
  case with parameters $\beta=1$ and $\beta=2$, respectively.  $\Gamma$-${\rm K}$-${\rm K'}$-$\Gamma$ is
  the high-symmetry path. The red, blue, and green solid lines correspond to the dispersions of the bands from
  the lowest to the highest, respectively. Other involved parameter is $\hbar\omega=9J$.}\label{f3}
\end{figure}

 \section{Chern number and edge state}
 From the SU(3) mapping, we know that the effective Hamiltonian shows a three-level system. There are three quasi-energy
 bands in the quasi-energy Brillouin zone $(-\omega/2,~\omega/2]$ (here $\hbar=1$), denoted by $E_{n}(\bm{k})$ with 
 $n=1, 2, 3$. The increasing $n$ corresponds to the $n$-th quasi-energy band arranged in an ascending order. For the $n$-th 
 band, its associated Chern number is \cite{TKNN,CN_cal_1,CN_def_1,CN_def_2,CN_cal_2}
 \begin{equation}\label{CN}
 C_{n}=\frac{1}{2\pi}\int_{FBZ} \Omega_{n}(\bm{k})d^2\bm{k},
 \end{equation}
 where the integration extends over the first Brillouin zone (FBZ) and the $\bm{\Omega}_{n}$ is the Berry curvature, which is defined
 in terms of the partial derivative of the eigenvector $\ket{\psi_{n}(\bm{k})}$ of $\hat{H}_{\rm eff}(\bm{k})$ as $\Omega_{n}(\bm{k})=i\left(\braket{\frac{\partial \psi_{n}(\bm{k})}{\partial k_{x}} | \frac{\partial \psi_{n}(\bm{k})}{\partial k_{y}}}- h.c.\right)$.

 Here, we investigate the topological properties and the band structures of the driven dice system both in the
 isotropic and the anisotropic case. Without loss of generality, we choose $J_{1}=0.5J$ to characterize 
 the anisotropic case. By employing the definition of the Chern number in Eq.~(\ref{CN}), the topological phase 
 diagram that contains the Chern number of the lowest quasi-energy band $C_{1}$ as the function of $\beta$ is 
 plotted in Fig.~\ref{f3}(a), where the red dots correspond to the isotropic case and blue dots correspond 
 to the anisotropic one. The Chern numbers of the middle band for the two cases are equal to zero, which are not 
 shown in the phase diagram. Alternatively, the Chern numbers can be calculated by the analytical method (see the 
 derivation in Appendix \ref{A3}), completely consistent with the numerical ones. Intuitively, without driving, namely $\beta=0$,
 the system is topological trivial with $C_{1}=0$ and system keeps topologically non-trivial once the driving is introduced. 
 Differently, there are large Chern numbers $C_{1}=2$ for the isotropic case while $C_{1}=1$ for the anisotropic one. 
 In fact, the Chern number of the static system is ill-defined because of the gapless dispersions of bands (see Figs.~\ref{f3}(b1)
 and \ref{f3}(b2)). $C_{1}=0$ is used to conveniently characterize the trivial and gapless case. $\Gamma$-$\rm{K}$-$\rm{K'}$-$\Gamma$ is
 the high-symmetry path where $\rm{K}$ and $\rm{K'}$ are the singularities \cite{CN_cal_1,CN_cal_2}. On the contrary, in the topologically non-trivial phase,
 the bands are gapped. For instance, in the isotropic case, as shown in Figs.~\ref{f3}(c1) ($\beta=1$) and \ref{f3}(d1) ($\beta=2$), three quasi-energy bands
 are separated by the gaps. Similar circumstance appears in the anisotropic case as well ($\beta=1$ in Fig.~\ref{f3}(c2) and $\beta=2$ in Fig.~\ref{f3}(d2)).
 Besides, we notice that there is a difference between the two cases in the topologically non-trivial phase. For the isotropic case, the middle quasi-energy
 band is fully a flat band (see Figs.~\ref{f3}(c1) and \ref{f3}(d1)), but middle quasi-energy band is obviously distorted at the high-symmetry
 points $\rm{K}$ and $\rm{K'}$ in the anisotropic case.

 \begin{figure}[htp]
  \centering
  \includegraphics[width=0.5\textwidth,height=0.45\textwidth]{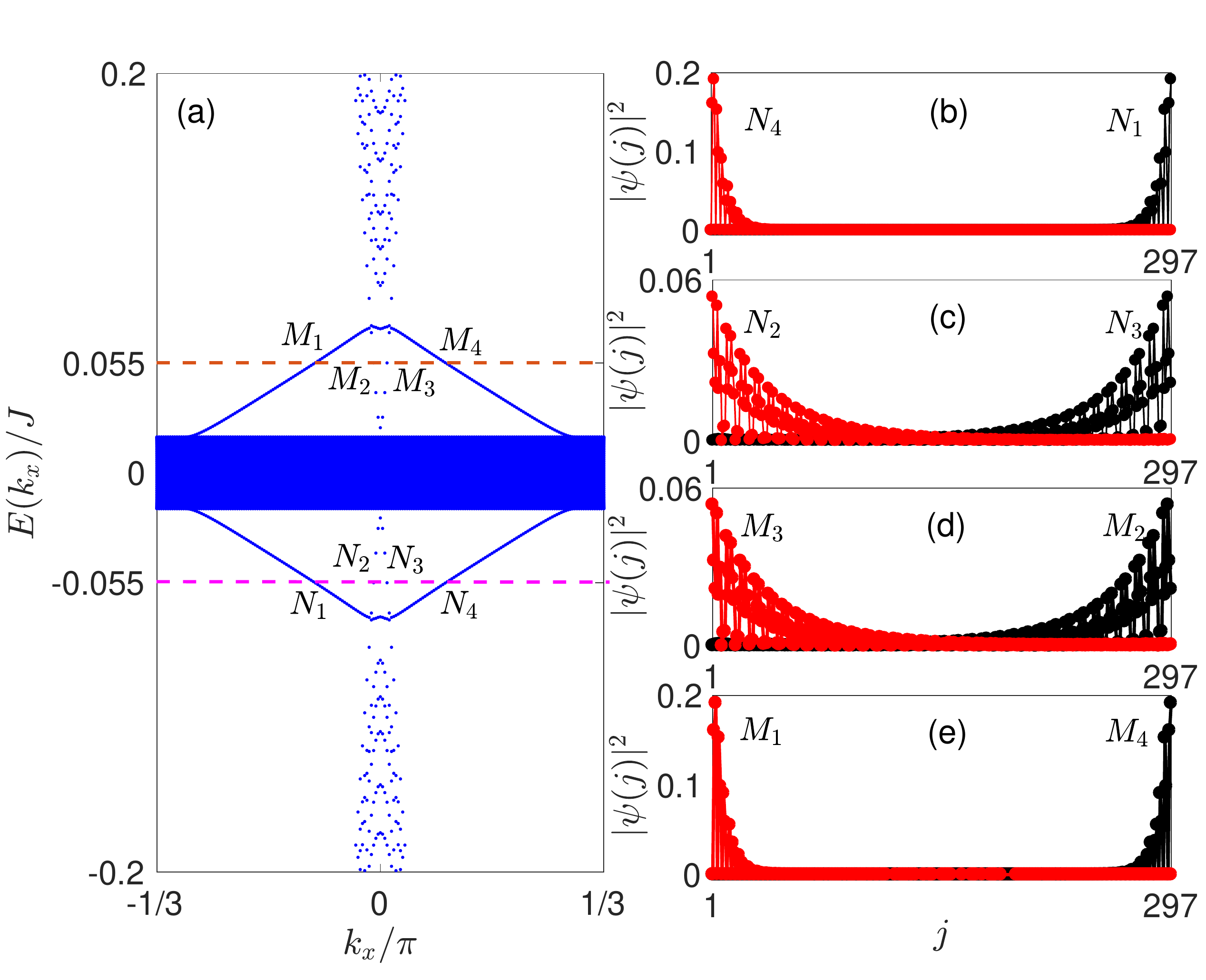}\\
  \caption{(Color Online) (a) Singly periodic quasi-energy spectrum $E(k_{x})$ of the isotropic case
  as a function of the quasi-momentum $k_{x}$. $N_{1}$, $N_{2}$, $N_{3}$, and $N_{4}$ are four edge
  modes chosen at $E(k_{x})\approx -0.055J$ (the magenta dashed line). $M_{1}$, $M_{2}$, $M_{3}$, and $M_{4}$
  are the ones chosen at $E(k_{x})\approx 0.055J$ (the orange dashed line). (b)-(e) Spatial distributions
  of these chosen edge modes. The modes with opposite quasi-momentum are symmetrically distributed
  at the edges of the dice geometry. The involved parameter is $\hbar\omega=9J$ and $J_1=0.5J$.
  }\label{f4}
\end{figure}

 Next, we select the isotropic case to discuss the correspondence between the Chern numbers and the edge modes
 according to the principle of the bulk-edge correspondence \cite{BECP} in such a Floquet system. In fact, the anisotropic
 case supports this principle as well (see Appendix \ref{A4}). After choosing a cylindrical dice geometry which preserves
 the periodicity in the $x$ direction but leaves it open in the $y$ direction (armchair edge), the singly periodic Bloch
 Hamiltonian $\hat{H}_{\rm{eff}}(k_{x})$ can be obtained by performing the partial Fourier transformation where $k_{x}$
 is the quantum number and $k_{x} \in \left[-\pi/3,\pi/3\right]$. In the numerical calculation, we consider that the
 super-cell contains total $N_s=297$ lattice sites and take $\beta=1$ and $J_{1}=J$. With these parameters,
 the singly periodic quasi-energy spectrum $E(k_{x})$ of the isotropic case is plotted in Fig.~\ref{f4}(a).
 $N_{1}$, $N_{2}$, $N_{3}$, and $N_{4}$ are four chosen edge modes within the lower bulk quasi-energy gap
 whose corresponding quasi-energies are $E(k_{x})\approx -0.055J$ (the dashed magenta line shows).
 $M_{1}$, $M_{2}$, $M_{3}$, and $M_{4}$ are another four edge modes within the upper bulk quasi-energy gap
 chosen at $E(k_{x})\approx 0.055J$ (the dashed orange line shows). The spatial distributions of these edge modes
 are plotted in Figs.~\ref{f4}(b)-\ref{f4}(e). Particularly, the red curves character the modes with positive group
 velocity (PGV) while the black ones character the modes with negative group velocity (NGV). Intuitively, the modes
 with opposite quasi-momentum are symmetrically distributed at the edges of the dice geometry. Without loss of generality,
 we select the modes localized at the $j=1$ side to analyze the bulk-edge correspondence. Since $C_{1}=2$, the modes
 $N_{2}$, $N_{4}$, $M_{1}$, and $M_{3}$ with PGV all carry the Chern number $C=1$. As Ref. \cite{BECP} tells, the
 Chern number of each band is the difference between the total Chern number carried by all the edge modes localized
 at one side above the band and the total Chern number carried by all the edge modes localized at the same side below
 the band. Therefore, we extract the Chern number of the flat middle band $C_{2}$ as $C_{2}=1+1-(1+1)=0$, which
 is in accord with our results.

 \section{Summary}
 In summary, the Floquet band engineering on the optical dice lattice has been well studied. Although the initial dice system possesses
 a gapless band structure, we uncover that the applied circular-frequency shaking will induce gapped quasi-energy bands and this non-trivial
 band characteristic persists within a large strength of the shaking force. Furthermore, after investigating the topological properties of the
 isotropic case and the anisotropic case of the driven system, we find that in the isotropic case, there exists a topological phase with Chern
 number $C_{1}=2$, higher than the one with $C_{1}=1$ in the anisotropic case. In the end, we discuss how to employ the associated edge
 modes to analyze the Chern number of quasi-energy bands within the framework of the principle of bulk-edge correspondence.
 Our detailed numerical and analytical calculations show that the idea of employing the circular-frequency shaking to induce the non-trivially
 topological characteristics of the optical dice lattice is theoretically feasible, and the topological phase transition can be achieved by
 independently tuning the hopping strength in one hopping direction. However, in the way to observe these theoretical predictions
 on a platform, we need to construct the light dice lattice, shake the light dice model, and then measure the topological phenomenon in experiment \cite{challenge_2}, which are the next interesting research topic.

 \begin{acknowledgments}
The authors acknowledge support from NSFC under
Grants No. 11835011 and No. 12174346. We
benefited greatly from discussions with Dr. Markus Schmitt and Dr. Pei Wang.
 \end{acknowledgments}

\begin{appendix}
\section{Derivation of $\hat{H}_{\rm rot}$}\label{A1}
 The $\hat{H}_{\rm rot}$ in Eq.~(3) can be expanded as
 \begin{equation}
 \begin{aligned}
 \hat{H}_{\rm rot}&=\hat{U}^\dag(t)\hat{H}_{\rm ini}\hat{U}(t)\\
 &=\sum_{\langle \bm{R}_j,\bm{B}_{j'}\rangle}J\left(\hat{U}^\dag(t)\hat{c}^\dag_{ \bm{R}_j}\hat{U}(t)\hat{U}^\dag(t)\hat{c}_{ \bm{B}_{j'}}\hat{U}(t)+H.c.\right)\\
 &+\sum_{\langle \bm{G}_{j}, \bm{R}_{j'}\rangle}J_1\left(\hat{U}^\dag(t)\hat{c}^\dag_{ \bm{G}_j}\hat{U}(t)\hat{U}^\dag(t)\hat{c}_{ \bm{R}_{j'}}\hat{U}(t)+H.c.\right),
 \end{aligned}
 \end{equation}
where $\hat{U}(t)=e^{\frac{i}{\hbar}\sum_{\bm{\alpha}_{j}}\int^{t}_{0}\left[\bm{\alpha}_{j}\cdot \bm{F}(t)\right]dt\cdot\hat{n}_{\bm{\alpha}_{j}}}$. 
Employing the Baker-Campbell-Hausdorff formula \cite{Eckardt,Floquet_1,Floquet_2},
 \begin{equation}
 \begin{aligned}
 e^{i\hat{X}}\hat{Y}e^{-i\hat{X}}&=\hat{Y}+i[\hat{X},\hat{Y}]+\frac{i^2}{2!}[\hat{X},[\hat{X},\hat{Y}]]\\
 &+\frac{i^3}{3!}[\hat{X},[\hat{X},[\hat{X},\hat{Y}]]]\ldots,
 \end{aligned}
 \end{equation}
we have
\begin{equation}
\hat{U}^\dag(t)\hat{c}^\dag_{\bm{\alpha}_{j}}\hat{U}(t)=e^{-\frac{i}{\hbar}\int^{t}_{0}\bm{F}(t)dt \cdot \bm{\alpha}_{j}}\hat{c}^{\dag}_{\bm{\alpha}_{j}},
\end{equation}
and
\begin{equation}
\hat{U}^\dag(t)\hat{c}_{\bm{\alpha'}_{j'}}\hat{U}(t)=e^{\frac{i}{\hbar}\int^{t}_{0}\bm{F}(t)dt \cdot \bm{\alpha'}_{j'}}\hat{c}_{\bm{\alpha'}_{j'}}.
\end{equation}

Therefore, the $\hat{H}_{\rm rot}$ is derived as
 \begin{equation}
 \begin{aligned}
 \hat{H}_{\rm rot}&=\sum_{\langle \bm{R}_j,\bm{B}_{j'}\rangle}J\left(e^{-i\frac{Fa}{\hbar\omega}\sin(\omega t+\theta^{\bm{R}_{j}}_{\bm{B}_{j'}})}\hat{c}^\dag_{\bm{R}_j}\hat{c}_{\bm{B}_{j'}}+H.c.\right)\\
 &+\sum_{\langle \bm{G}_{j},\bm{R}_{j'}\rangle}J_1\left(e^{-i\frac{Fa}{\hbar\omega}\sin(\omega t+\theta^{\bm{G}_{j}}_{\bm{R}_{j'}})}\hat{c}^\dag_{\bm{G}_j}\hat{c}_{\bm{R}_{j'}}+H.c.\right),
 \end{aligned}
 \end{equation}
 where $\theta^{\bm{\alpha}_{j}}_{\bm{\alpha}'_{j'}}$ the direction angle from site $\bm{\alpha}'_{j'}$ to its neighbor $\bm{\alpha}_{j}$.

\section{Derivation of $\hat{\mathcal{H}}_{m}$}\label{A2}
The hopping strength in $\hat{H}_{\rm rot}$ can be rewritten as
\begin{equation}
 e^{-i\beta\sin(\omega t+\theta^{\bm{\alpha}'_{j'}}_{\bm{\alpha}_{j}})}=exp\left[\beta\frac{e^{-i(\omega t+\theta^{\alpha_{j}}_{\alpha'_{j'}})}-e^{i(\omega t+\theta^{\alpha_{j}}_{\alpha'_{j'}})}}{2}\right].
\end{equation}

Employing the Jacobi-Anger expansion
 \begin{equation}
 \exp\left[\xi\frac{x-x^{-1}}{2}\right]=\sum^{\infty}_{\ell=-\infty}\mathcal{J}_{\ell}(\xi)x^{\ell},
 \end{equation}
 $\hat{H}_{\rm rot}$ is written as
 \begin{equation}
 \begin{aligned}
  \hat{H}_{\rm rot}&=\sum_{\langle \bm{R}_j,\bm{B}_{j'}\rangle}J\left(\sum^{\infty}_{\ell=-\infty}\mathcal{J}_{\ell}(\beta)e^{-i\ell(\omega t+\theta^{\bm{R}_{j}}_{\bm{B}_{j'}})}\hat{c}^\dag_{\bm{R}_j}\hat{c}_{\bm{B}_{j'}}+H.c.\right)\\
 +&\sum_{\langle \bm{G}_{j},\bm{R}_{j'}\rangle}J_1\left(\sum^{\infty}_{\ell=-\infty}\mathcal{J}_{\ell}(\beta)e^{-i\ell(\omega t+\theta^{\bm{G}_{j}}_{\bm{R}_{j'}})}\hat{c}^\dag_{\bm{G}_j}\hat{c}_{\bm{R}_{j'}}+H.c.\right).
 \end{aligned}
 \end{equation}

Noticing that $\hat{H}_{\rm rot}$ is time-periodic, then it can be expanded as
 \begin{equation}
 \hat{H}_{\rm rot}=\sum_{m=-\infty}^{\infty}\hat{\mathcal{H}}_{m}e^{im\omega t},
 \end{equation}
where $\hat{\mathcal{H}}_{m}$ is the $m$-th Fourier component of $\hat{H}_{\rm rot}$, from which we
derive the $\hat{\mathcal{H}}_{m}$ as
\begin{equation}
 \begin{aligned}
 &\hat{\mathcal{H}}_{m}=\frac{\omega}{2\pi} \int^{\frac{2\pi}{\omega}}_{0}\hat{H}_{\rm rot} e^{-im\omega t} dt\\
 &=\sum_{\langle \bm{R}_j,\bm{B}_{j'}\rangle}J\left(\mathcal{J}_{-m}(\beta)e^{im\theta^{\bm{R}_{j}}_{\bm{B}_{j'}}}\hat{c}^\dag_{\bm{R}_j}\hat{c}_{\bm{B}_{j'}}
 +\mathcal{J}_{m}(\beta)e^{im\theta^{\bm{R}_{j}}_{\bm{B}_{j'}}}\hat{c}^\dag_{\bm{B}_{j'}}\hat{c}_{\bm{R}_{j}}\right)\\
 &+\sum_{\langle \bm{G}_{j},\bm{R}_{j'}\rangle}J_1\left(\mathcal{J}_{-m}(\beta)e^{im\theta^{G_{j}}_{R_{j'}}}\hat{c}^\dag_{\bm{G}_j}\hat{c}_{\bm{R}_{j'}}
 +\mathcal{J}_{m}(\beta)e^{im\theta^{\bm{G}_{j}}_{\bm{R}_{j'}}}\hat{c}^\dag_{\bm{R}_{j'}}\hat{c}_{\bm{G}_{j}}\right).
 \end{aligned}
 \end{equation}

\section{Derivation of the Chern number}\label{A3}
 We derive the analytical Chern number of the Hamiltonian presented in Eq.~(\ref{CN}).
 In principle, all the eigenenergies and wavefunctions can be exactly solved, by which we can derive the Berry connection or
 the Berry curvature and then calculate the Chern number of each band after performing an integration \cite{TKNN}.
 However, the directly obtained eigenvalues and wavefunctions are rather complicated, and are not convenient for us
 to derive the Berry connection or the Berry curvature directly. Therefore, we adopt an unconventional strategy to calculate
 the Chen number.

 In the derivation, we suppose that each eigenvalue has a clear expression in advance, but we do not know which band it belongs to.
 For a given eigenenergy $\lambda_{s}(\bm k)$ ($s=1,2,3$), its corresponding wavefunction $\ket{u_{s}(\bm k)}$ is given as
 \begin{widetext}
 \begin{equation}
 \ket{u_{s}(\bm k)}=\left(
  \begin{array}{c}
 \frac{(\lambda_{s}(\bm k)+d_3-d_8)(\lambda_{s}(\bm k)+2d_8)}{\sqrt{(\lambda_{s}(\bm k)+2d_8)^2(\lambda_{s}(\bm k)+d_3-d_8)^2+(d^2_1+d^2_2)(\lambda_{s}(\bm k)+2d_8)^2+(d^2_4+d^2_5)(\lambda_{s}(\bm k)+d_3-d_8)^2}}\\
 ~\\
 \frac{(d_{1}+id_2)(\lambda_{s}(\bm k)+2d_8)}{\sqrt{(\lambda_{s}(\bm k)+2d_8)^2(\lambda_{s}(\bm k)+d_3-d_8)^2+(d^2_1+d^2_2)(\lambda_{s}(\bm k)+2d_8)^2+(d^2_4+d^2_5)(\lambda_{s}(\bm k)+d_3-d_8)^2}}\\
 ~\\
 \frac{(d_4+id_5)(\lambda_{s}(\bm k)+d_3-d_8)}{\sqrt{(\lambda_{s}(\bm k)+2d_8)^2(\lambda_{s}(\bm k)+d_3-d_8)^2+(d^2_1+d^2_2)(\lambda_{s}+2d_8)^2+(d^2_4+d^2_5)(\lambda_{s}(\bm k)+d_3-d_8)^2}}\\
 \end{array}
 \right).
 \end{equation}
 \end{widetext}
 With $\ket{\psi_{n}(\bm k)}$, we derive the Berry connection as
 \begin{widetext}
 \begin{equation}\label{A}
 \begin{aligned}
 \vec{A}&=-i\braket{u_{s}(\bm k)| \nabla_{\bm k}| u_{s}(\bm k)}\\
 &=\frac{(d_1\nabla_{\bm k}{d_2}-d_2\nabla_{\bm k}d_1)(\lambda_{s}(\bm k)+2d_8)^2}{(\lambda_{s}(\bm k)+2d_8)^2(\lambda_{s}(\bm k)+d_3-d_8)^2+(d^2_1+d^2_2)(\lambda_{s}(\bm k)+2d_8)^2+(d^2_4+d^2_5)(\lambda_{s}(\bm k)+d_3-d_8)^2} \\
 &+\frac{(d_4\nabla_{\bm k}d_5-d_5\nabla_{\bm k}d_4)(\lambda_{s}(\bm k)+d_3-d_8)^2}{(\lambda_{s}(\bm k)+2d_8)^2(\lambda_{s}(\bm k)+d_3-d_8)^2+(d^2_1+d^2_2)(\lambda_{s}(\bm k)+2d_8)^2+(d^2_4+d^2_5)(\lambda_{s}(\bm k)+d_3-d_8)^2},
 \end{aligned}
 \end{equation}
 \end{widetext}

 According to the generation of the TKNN theory in the three-band system \cite{CN_cal_1} and the one in the two-band system \cite{CN_cal_2},
 we know that the Chern number of the band is contributed by the singularity $\bm{q}$, at which the Berry connection $\vec{A}$ is singular.
 By analyzing the expression of $\vec{A}$ in Eq.~(\ref{A}), we extract that there are two types of singularities $\bm{q}^{(1)}$ and $\bm{q}^{(2)}$
 in such a system. The first-type singularity $\bm{q}^{(1)}$ makes
 \begin{equation}
 d_{1}=d_{2}=0,
 \end{equation}
 and contributes non-zero Chern numbers to the band with $\lambda_{s}(\bm{k}=\bm{q}^{(1)})=-d_3+d_8$. The second one $\bm{q}^{(2)}$ makes
\begin{equation}
d_{4}=d_{5}=0,
\end{equation}
and contributes non-zero Chern numbers to the band with $\lambda_{s}(\bm{k}=\bm{q}^{(2)})=-2d_8$.

We first discuss the first type case. In a concrete system, if there are more than one singularity satisfying
the first-type of singularity condition, around the infinitesimal neighborhood of each first-type singularity $\bm{q}^{(1)}_{j}$, the corresponding
matrix elements can be expanded as
\begin{equation}
\begin{aligned}
d^{\bm{q}^{(1)}_{j}}_1&=a^{\bm{q}^{(1)}_{j}}_{1x}\Delta k_{x}+a^{\bm{q}^{(1)}_{j}}_{1y}\Delta k_{y} +\mathcal{O}(\Delta \bm{k}^2),\\
d^{\bm{q}^{(1)}_{j}}_2&=a^{\bm{q}^{(1)}_{j}}_{2x}\Delta k_{x}+a^{\bm{q}^{(1)}_{j}}_{2y}\Delta k_{y} +\mathcal{O}(\Delta \bm{k}^2).
\end{aligned}
\end{equation}
 Then, the Chern number contributed by $\bm{q}^{(1)}_{j}$ is
\begin{equation}\label{cq1}
\begin{aligned}
C_{\bm{q}^{(1)}_{j}}&=\frac{1}{2\pi}\oint_{\partial{\bm{q}^{(1)}_{j}}}\vec{A}_{\bm{q}^{(1)}_{j}}\cdot d\bm{k}\\
&=sgn(a^{\bm{q}^{(1)}_{j}}_{1x}a^{\bm{q}^{(1)}_{j}}_{2y}-a^{\bm{q}^{(1)}_{j}}_{2x}a^{\bm{q}^{(1)}_{j}}_{1y}).
\end{aligned}
\end{equation}

 For the second-type case, there may exist more than one singularity satisfying the singularity condition as well.
 Around the infinitesimal neighborhood of each second-type singularity $\bm{q}^{(2)}_{j}$, the corresponding
 $d^{\bm{q}^{(2)}_{j}}_{1}$ and $d^{\bm{q}^{(2)}_{j}}_{2}$ can be expanded as the similar form
 \begin{equation}
 \begin{aligned}
 d^{\bm{q}^{(2)}_{j}}_1&=a^{\bm{q}^{(2)}_{j}}_{4x}\Delta k_{x}+a^{\bm{q}^{(2)}_{j}}_{4y}\Delta k_{y} +\mathcal{O}(\Delta \bm{k}^2),\\
 d^{\bm{q}^{(2)}_{j}}_2&=a^{\bm{q}^{(2)}_{j}}_{5x}\Delta k_{x}+a^{\bm{q}^{(2)}_{j}}_{5y}\Delta k_{y} +\mathcal{O}(\Delta \bm{k}^2).
 \end{aligned}
 \end{equation}
 Then, the Chern number contributed by $\bm{q}^{(2)}_{j}$ is
 \begin{equation}\label{cq2}
 \begin{aligned}
 C_{\bm{q}^{(2)}_{j}}&=\frac{1}{2\pi}\oint_{\partial{\bm{q}^{(2)}_{j}}}\vec{A}_{\bm{q}^{(2)}_{j}}\cdot d\bm{k}\\
 &=sgn(a^{\bm{q}^{(2)}_{j}}_{4x}a^{\bm{q}^{(2)}_{j}}_{5y}-a^{\bm{q}^{(2)}_{j}}_{5x}a^{\bm{q}^{(2)}_{j}}_{4y}).
 \end{aligned}
 \end{equation}

 Up to now, we have known the types of singularities in this generalized system and the expressions of
 the Chern numbers they contribute. Moreover, from their expressions, we know that $C_{\bm{q}^{(1)}_{j}}$ and
 $C_{\bm{q}^{(2)}_{j}}$ only depend on the expansion coefficients while having nothing to with whether the system
 is in the isotropic case or the anisotropic one. Nevertheless, two key problems remain to be solved.
 The first one is that which bands $\lambda_{s}(\bm{q}^{(1)}_{j})$ and $\lambda_{s}(\bm{q}^{(2)}_{j})$ correspond to.
 The second one is the sum of Chern numbers contributed by the two kinds of singularities to each band.
 To answer the questions, it is necessary to analyze the eigenenergies at the singularities.

 Around the first-type singularity $\bm{q}^{(1)}_{j}$, the Hamiltonian can be reexpressed as
 \begin{equation}
 \begin{aligned}
 \hat{H}^{\bm{q}^{(1)}_{j}}_{\rm eff}&=\hat{H}_{0}(\bm{k}=\bm{q}^{(1)}_{j})+\hat{H}'_{\bm{q}^{(1)}_{j}}\\
 &=\left(
 \begin{array}{ccc}
 d_3+d_8 & 0 & d_4-id_5 \\
 0 & -d_3+d_8 & 0 \\
 d_4+id_5 & 0 & -2d_8
 \end{array}
 \right)\\
& +\left(
 \begin{array}{ccc}
 0 & d^{\bm{q}^{(1)}_{j}}_1-id^{\bm{q}^{(1)}_{j}}_2 & 0  \\
 d^{\bm{q}^{(1)}_{j}}_1+id^{\bm{q}^{(1)}_{j}}_2 & 0 & 0 \\
 0 & 0 & 0
 \end{array}
 \right),
 \end{aligned}
 \end{equation}
 where $\hat{H}'_{\bm{q}^{(1)}_{j}}$ is regarded as the perturbation term. Under the second-order perturbation
 approximation, the eigenenergies around $\bm{q}^{(1)}_{j}$ are
 \begin{equation}
 \begin{aligned}
 \lambda_{1}(\bm{q}^{(1)}_{j})&=-d_3+d_8, \\
 \lambda_{2}(\bm{q}^{(1)}_{j})&=\frac{-d_3+d_8}{2}-\sqrt{(\frac{d_3+3d_8}{2})^2+d^2_4+d^2_5},\\
 \lambda_{3}(\bm{q}^{(1)}_{j})&=\frac{-d_3+d_8}{2}+\sqrt{(\frac{d_3+3d_8}{2})^2+d^2_4+d^2_5}.
 \end{aligned}
 \end{equation}

 We can determine the Chern number of the three bands just by comparing $\lambda_{1}(\bm{q}^{(1)}_{j})$
 with $\lambda_{2}(\bm{q}^{(1)}_{j})$ and $\lambda_{3}(\bm{q}^{(1)}_{j})$. For instance, if $\lambda_{1}(\bm{q}^{(1)}_{j})$
 is the smallest one among the three eigenenergies, i.e., $\lambda_{1}(\bm{q}^{(1)}_{j}) \equiv E_{1}(\bm{q}^{(1)}_{j})$,
 then the lowest band $E_{1}$ has the non-zero Chern number $C_{\bm{q}^{(1)}_{j}}$, while the Chern number of other two
 bands are both equal to zero.

 In the same way, we reexpress the Hamiltonian around the second-type singularity $\bm{q}^{(2)}_{j}$ as
 \begin{equation}
 \begin{aligned}
 \hat{H}^{\bm{q}^{(2)}_{j}}_{\rm eff}&=\hat{H}_{0}(\bm{k}=\bm{q}^{(2)}_{j})+\hat{H}'_{\bm{q}^{(2)}_{j}}\\
 &=\left(
 \begin{array}{ccc}
 d_3+d_8 & d_1-id_2 & 0 \\
 d_1+id_2 & -d_3+d_8 & 0 \\
 0 & 0 & -2d_8
 \end{array}
 \right)\\
 &+\left(
 \begin{array}{ccc}
 0 & 0 & d^{\bm{q}^{(2)}_{j}}_4-id^{\bm{q}^{(2)}_{j}}_5  \\
 0 & 0 & 0 \\
 d^{\bm{q}^{(2)}_{j}}_4+id^{\bm{q}^{(2)}_{j}}_5 & 0 & 0
 \end{array}
 \right),
 \end{aligned}
 \end{equation}
 where $\hat{H}'_{\bm{q}^{(2)}_{j}}$ is regarded as the perturbation term. Under the second-order perturbation
 approximation, the eigenenergies around $\bm{q}^{(2)}_{j}$ are
 \begin{equation}
 \begin{aligned}
 \lambda_{1}({\bm{q}^{(2)}_{j}})&=-2d_8, \\
 \lambda_{2}({\bm{q}^{(2)}_{j}})&=d_8-\sqrt{d^2_1+d^2_2+d^2_3},\\
 \lambda_{3}({\bm{q}^{(2)}_{j}})&=d_8+\sqrt{d^2_1+d^2_4+d^2_5}.
 \end{aligned}
 \end{equation}

 Following the same analysis method as the first-type case, by comparing $\lambda_{1}({\bm{q}^{(2)}_{j}})$ with
 $\lambda_{2}({\bm{q}^{(2)}_{j}})$ and $\lambda_{3}({\bm{q}^{(2)}_{j}})$, we can determine which band
 the $\lambda_{1}({\bm{q}^{(2)}_{j}})$ corresponds to. If $\lambda_{1}({\bm{q}^{(2)}_{j}})$ is the largest one
 among the three eigenenergies, then the highest band $E_{3}$ has a non-zero Chern number $C_{\bm{q}^{(2)}_{j}}$.
 Otherwise, the middle band $E_{2}$ or the lowest band $E_{1}$ has a non-zero Chern number $C_{\bm{q}^{(2)}_{j}}$.
 Based on the above analysis, we conclude that the Chern number of a concrete band is the summation of the
 Chern numbers contributed by all singularities to the band.

In the following, we choose the isotropic case with $J_{1}=J$ and $\beta=1$ and the anisotropic case with
$J_{1}=0.5J$ and $\beta=2$ as two examples, and then calculate the Chern numbers of the two examples
by this analytical method.  After comparing the four matrix elements $d_{1}$, $d_{2}$, $d_{4}$, and $d_{5}$ in Eq.~(\ref{elements}),
we find that the singularities satisfying the first-type singularity condition satisfy the second-type singularity condition as well.
To obtain the Chern number of the special case, we just have to substitute the expansion
coefficients of the four matrix elements into the definitions of the first-type Chern number and the
second-type one, respectively. Then we calculate the Chern number of each band according to the
above-mentioned summation rule.

From the singularity condition, we extract two singularities simultaneously satisfying the first-type and second-type conditions.
One is $\bm{q}^{(1)}_{1} \left(\bm{q}^{(2)}_{1}\right)=\rm{\bm{K}}\equiv\left(\frac{4\pi}{3\sqrt{3}},0\right)$ and the other is
$\bm{q}^{(1)}_{2} \left(\bm{q}^{(2)}_{2}\right)=\rm{\bm{K}}'=\left(\frac{2\pi}{3\sqrt{3}},\frac{2\pi}{3}\right)$. For the first-type
singularity case, the expansion coefficients are
\begin{equation}
\begin{aligned}
&a^{\bm{q}^{(1)}_{1}}_{1x}=-\frac{3}{2}t_{rb},~a^{\bm{q}^{(1)}_{1}}_{1y}=0,\\
&a^{\bm{q}^{(1)}_{1}}_{2x}=0,~a^{\bm{q}^{(1)}_{1}}_{2y}=-\frac{3}{2}t_{rb},\\
&a^{\bm{q}^{(1)}_{2}}_{1x}=-\frac{3}{4}t_{rb},~a^{\bm{q}^{(1)}_{2}}_{1y}=-\frac{3\sqrt{3}}{4}t_{rb},\\
&a^{\bm{q}^{(1)}_{2}}_{2x}=-\frac{3\sqrt{3}}{4}t_{rb},~a^{\bm{q}^{(1)}_{2}}_{2y}=\frac{3}{4}t_{rb}.
\end{aligned}
\end{equation}
From Fig.~\ref{f2}, we know that the parameter $t_{rb}$ and $t_{gr}$ are indeed positive numbers either in the isotropic case or
in the anisotropic case. Substituting these expansion coefficients into the definition of $C_{\bm{q}^{(1)}_{j}}$ in Eq.~(\ref{cq1}),
we have
\begin{equation}
C_{\bm{q}^{(1)}_{1}}=1,~C_{\bm{q}^{(1)}_{2}}=-1.
\end{equation}

For the second-type singularity case, the expansion coefficients are
\begin{equation}
\begin{aligned}
&a^{\bm{q}^{(2)}_{1}}_{4x}=-\frac{3}{2}t_{gr},~a^{\bm{q}^{(2)}_{1}}_{4y}=0,\\
&a^{\bm{q}^{(2)}_{1}}_{5x}=0,~a^{\bm{q}^{(2)}_{1}}_{5y}=\frac{3}{2}t_{gr},\\
&a^{\bm{q}^{(2)}_{2}}_{4x}=-\frac{3}{4}t_{gr},~a^{\bm{q}^{(2)}_{2}}_{4y}=-\frac{3\sqrt{3}}{4}t_{gr},\\
&a^{\bm{q}^{(2)}_{2}}_{5x}=\frac{3\sqrt{3}}{4}t_{gr},~a^{\bm{q}^{(2)}_{2}}_{5y}=-\frac{3}{4}t_{gr}.
\end{aligned}
\end{equation}
Substituting the expansion coefficients into the definition of $C_{\bm{q}^{(2)}_{j}}$ in Eq.~(\ref{cq2}), we have
\begin{equation}
C_{\bm{q}^{(2)}_{1}}=-1,~C_{\bm{q}^{(2)}_{2}}=1.
\end{equation}

Next, we analyze that which band $C_{\bm{q}^{(1)}_{j}}$ or $C_{\bm{q}^{(2)}_{j}}$ corresponds to. At $\bm{q}^{(1)}_{j}$,
the eigenenergies are
\begin{equation}
\begin{aligned}
\lambda_{1}(\bm{q}^{(1)}_{j})&=\sum_{s=1,2,3} 2|t_{bb}|\sin\left(\bm{k}\cdot \bm{q}^{(1)}_{j}\right), \\
\lambda_{2}(\bm{q}^{(1)}_{j})&=\sum_{s=1,2,3}-2|t_{rr}| \sin\left(\bm{k}\cdot \bm{q}^{(1)}_{j}\right), \\
\lambda_{3}(\bm{q}^{(1)}_{j})&= \sum_{s=1,2,3}-2|t_{gg}|\sin\left(\bm{k}\cdot \bm{q}^{(1)}_{j}\right). \\
\end{aligned}
\end{equation}
After comparing the three eigenenergies, we find that $C_{\bm{q}^{(1)}_{1}}=1$ corresponds to the lowest band $E_{1}$
and $C_{\bm{q}^{(1)}_{2}}=1$ corresponds to the highest band $E_{3}$ both in the isotropic and anisotropic cases.

At $\bm{q}^{(2)}_{j}$, the eigenenergies are
\begin{equation}
\begin{aligned}
\lambda_{1}(\bm{q}^{(2)}_{j})&=\sum_{s=1,2,3} -2|t_{gg}|\sin\left(\bm{k}\cdot \bm{q}^{(2)}_{j}\right), \\
\lambda_{2}(\bm{q}^{(2)}_{j})&=\sum_{s=1,2,3} 2|t_{bb}| \sin\left(\bm{k}\cdot \bm{q}^{(2)}_{j}\right), \\
\lambda_{3}(\bm{q}^{(2)}_{j})&= \sum_{s=1,2,3} -2|t_{rr}|\sin\left(\bm{k}\cdot \bm{q}^{(2)}_{j}\right). \\
\end{aligned}
\end{equation}
Similarly, by comparing the three eigenenergies, we find that in the isotropic case, $C_{\bm{q}^{(2)}_{1}}=-1$
corresponds to the highest band $E_{3}$ and $C_{\bm{q}^{(2)}_{1}}=1$ corresponds to the lowest band $E_{1}$,
whereas in the anisotropic case, both $C_{\bm{q}^{(2)}_{1}}=-1$ and $C_{\bm{q}^{(2)}_{2}}=1$ correspond to the
middle band $E_{2}$. Synthesizing the above analysis, we have: In the isotropic case, the Chern numbers are $C_{1}=1+1=2$,
$C_{2}=0$, and $C_{3}=-2$, respectively; in the anisotropic case, the Chern numbers are $C_{1}=1$, $C_{2}=0$, and
$C_{3}=-1$, respectively.

 \begin{figure}[htp]
  \centering
  \includegraphics[width=0.5\textwidth]{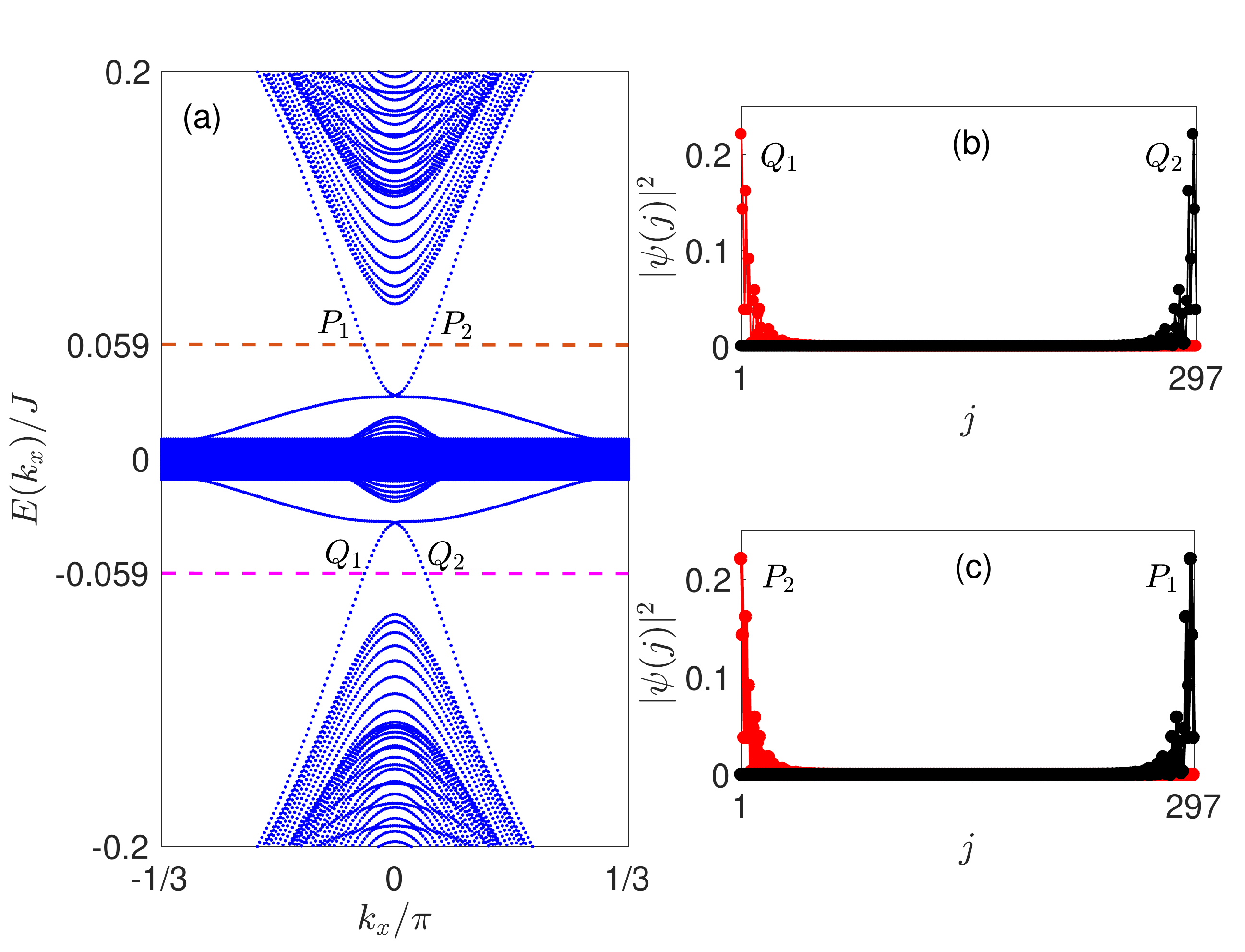}\\
  \caption{(Color Online) (a) Singly periodic quasi-energy spectrum $E(k_{x})$ of the anisotropic case
  as a function of the quasi-momentum $k_{x}$. $Q_{1}$ and $Q_{2}$ are a pair of edge modes chosen
  at $E(k_{x})\approx -0.059J$ (the magenta dashed line). $P_{1}$ and $P_{2}$ are another pair of
  edge modes chosen at $E(k_{x})\approx 0.059J$ (the orange dashed line). Panels (b) and (c) present
  spatial distributions of these chosen edge modes. The modes with opposite quasi-momentum are symmetrically
  distributed at the edges of the dice geometry.
  }\label{f5}
\end{figure}

\section{The bulk-edge correspondence in the anisotropic case}\label{A4}
Still considering the armchair dice geometry and taking $N_{s}=297$, the singly periodic quasi-energy
spectrum $E(k_{x})$ of the anisotropic case ($\hbar\omega=9J$, $\beta=2$ and $J_1=0.5J$) is plotted in Fig.~\ref{f5}(a).
$Q_{1}$ and $Q_{2}$ are a pair of chosen edge mode with opposite quasi-momentum $k_{x}$
within the lower bulk quasi-energy gap, whose corresponding quasi-energies are $E(k_{x})\approx -0.059J$.
$P_{1}$ and $P_{2}$ are another pair of chosen edge modes with opposite $k_{x}$ within the
upper bulk gap. The corresponding quasi-energies are $E(k_{x})\approx 0.059J$. Figures \ref{f5}(b) and \ref{f5}(c) 
present the spatial distributions of these chosen edge modes. It is readily seen that the modes with opposite 
quasi-momentum are symmetrically distributed at the edges of the dice geometry. We analyze the bulk-edge correspondence
by selecting the modes localized at the $j=1$ side. As discussed in the isotropic case, the modes
$Q_{1}$ and $P_{2}$ with PGV both correspond to the Chern number $C=1$. Therefore, we know that the
Chern number of the lowest quasi-energy band $C_{1}$ is $C_{1}=1$ and the Chern number of the middle
quasi-energy band $C_{2}$ is $C_{2}=1-1=0$, which are the same as the numerical and analytical results.
\end{appendix}

\bibliography{references}
\end{document}